# SharedDB: Killing One Thousand Queries With One Stone

Georgios Giannikis  Gustavo Alonso  Donald Kossmann

Systems Group, Department of Computer Science, ETH Zurich, Switzerland
{giannikg,alonso,kossmann}@inf.ethz.ch

## ABSTRACT

Traditional database systems are built around the query-at-a-time model. This approach tries to optimize performance in a best-effort way. Unfortunately, best effort is not good enough for many modern applications. These applications require response time guarantees in high load situations. This paper describes the design of a new database architecture that is based on batching queries and shared computation across possibly hundreds of concurrent queries and updates. Performance experiments with the TPC-W benchmark show that the performance of our implementation, SharedDB, is indeed robust across a wide range of dynamic workloads.

## 1. INTRODUCTION

Over the last decades, tremendous efforts have been invested into query optimization with the goal of achieving the best possible performance for each individual query in a query-at-a-time processing model. To this end, sophisticated query compile-time techniques (e.g., cost-based optimization [24]) and many kinds of database operator implementations (e.g., [26]) have been developed. While highly effective, unfortunately, all these approaches are not sufficient to meet common requirements of modern database applications if used in a query-at-a-time model. Often, modern applications need to meet service-level agreements (SLAs) that involve a maximum response time for, say, 99 percent of the queries. Furthermore, such SLAs may specify isolation levels and/or data freshness guarantees for workloads that involve queries and updates. The query-at-a-time processing model is not good to meet such SLAs because it may result in resource contention and interference in high load situations.

This paper describes a system, SharedDB, specifically designed to meet SLAs in high load situations for complex and highly dynamic workloads. SharedDB is based on a new processing model that batches queries and updates in order to share computation across these queries and updates. The key idea of SharedDB can be best described using an example involving two queries. The first query asks for all orders of German customers. The second query asks for all orders of Swiss customers in the Year 2011. SharedDB executes these two queries in a single customer-order join operation joining *all* orders with the union of German and Swiss customers and routing the results of the big join to the corresponding queries.

At a first glance, SharedDB looks like a bad idea. In the example, SharedDB performs extra work when compared to a traditional query-at-a-time approach. Specifically, SharedDB compares Swiss customers and orders of the Year 2010 as part of its join; such a combination is never considered when processing the two queries individually and aggressively pushing down predicates below the joins. Because of this, SharedDB is likely to perform poorly in low-throughput situations. SharedDB, however, was designed to handle high throughput with response time guarantees. If hundreds of concurrent queries involving a customer-order join need to be processed, then there will be a significant overlap between the sets of customers and orders relevant for these queries. The more queries, the higher the overlap, even if the queries involve totally different kinds of predicates on the Customer and Order tables. Furthermore, SharedDB defines an upper bound for the amount of work that needs to be carried out for a set of concurrent customer-order join queries. In the worst case, the whole customer table needs to be joined with the whole order table, independently of the number of concurrent queries processed. In contrast, the effort of a traditional, query-at-a-time system grows linearly with the number of queries. This way, SharedDB is able to maintain response time guarantees in high-throughput situations.

SharedDB adopts many ideas that were developed in the context of multi-query optimization and data stream processing. In particular, SharedDB adopts some of the ideas developed as part of QPipe [15], CJoin [3, 4] and DataPath [1]. These systems, however, are only effective for certain kinds of queries and, thus, their application is limited to OLAP workloads with complex queries. In contrast, this paper will show that the design principles of SharedDB are general and can be applied to any kind of query and update. As a result, SharedDB is able to process OLTP workloads in addition to OLAP and mixed workloads. It is this generality and its ability to meet SLAs in high load situations that distinguishes SharedDB from its closest competitors. Technically, it is the batch-oriented query processing model with a new way to share computation that makes SharedDB unique.

We have carried out comprehensive experiments using the TPC-W benchmark, thereby comparing the performance of SharedDB with that of MySQL and a top-of-the-line commercial relational database system. The results show that SharedDB is able to sustain twice the throughput of the top-of-the-line database system and almost eight times the throughput of MySQL. These results are surprising since the traditional database systems are much more mature and were specifically tuned to achieve highest possible throughput for benchmarks such as TPC-W. Furthermore, SharedDB does not apply any of the recently developed optimizations to achieve particularly good performance for OLTP workloads (e.g., [27]). We could not carry out experiments with QPipe, CJoin, or DataPath because these systems are not publicly available. Furthermore, it is not clear if and how the techniques used in these systems can be applied to transactional workloads such as TPC-W. Nevertheless, this paper contains a careful discussion of the differences and the advantages of SharedDB as compared to these systems.

The remainder of this paper is organized as follows: Section 2 revisits related work. Section 3 presents the main ideas of SharedDB and discusses in more detail the differences to other related approaches to share computation. Section 4 gives implementation details. Section 5 discusses the results of performance experiments. Section 6 contains conclusions and avenues for future work.

## 2. RELATED WORK

As mentioned in the introduction, SharedDB is based on shared computation that has recently been studied in several studies to optimize OLAP workloads. The idea of shared computation was first devised in the Eighties in the context of multi-query optimization (MQO) [11, 25]. The key idea of MQO is to detect common subexpressions in concurrently executed queries and to evaluate these common subexpressions only once. One problem of MQO is its limited applicability: In many workloads (in particular, transactional workloads such as TPC-W), there are not many opportunities to factor out common subexpressions. Another problem of classic MQO is that it is quite costly to detect common subexpressions. For these two reasons, MQO has been mostly applied to complex OLAP queries that involve the processing of large amounts of data and involve large common subexpressions (e.g., the scan of a fact table in a data warehouse) so that the investment of detecting common subexpressions is offset by the benefits of shared computation. As explained in the example of the introduction, SharedDB is able to carry out shared computation without common subexpression detection and is, thus, applicable to OLTP and mixed workloads.

Another problem of MQO is the *synchronization* of the execution of queries with common subexpressions when queries are submitted at different moments in time. To this end, the QPipe system developed a new query processing model that allows to exploit MQO if the queries were executed within a certain time frame [15]. The big advantage of this approach is that sharing does not slow down queries if the queries arrive at different points in time. Another contribution of the QPipe project is to differentiate between *data sharing* and *work sharing* and to provide a comprehensive taxonomy on the different forms of sharing for different kinds of database operators. Opportunities for data sharing arise in scan operations through base data; i.e., shared scans. Opportunities for work sharing in QPipe, however, only arise in the presence of common subexpressions. In order to unfold its full potential, therefore, QPipe also relies on common subexpression detection and has, thus, only been studied for OLAP workloads, just as classic MQO techniques.

Based on the QPipe work, two recent systems have exploited work sharing without the detection of common subexpressions: CJoin [3, 4] and DataPath [1]. SharedDB falls into this same class of systems so that CJoin and DataPath can be viewed as SharedDB's closest competitors. Indeed, there are a number of similarities in the design, but there are also fundamental differences. At the core of CJoin and DataPath are special, dedicated join algorithms in order to facilitate shared computation in a *pipelined* query execution model à la QPipe. In contrast, SharedDB is based on a *batched* query execution model. Furthermore, SharedDB uses standard query processing techniques such as index nested-loops, hashing and sorting for any kind of operator of the relational algebra (e.g., joins, grouping, ranking, and sorting) whereas CJoin and DataPath are limited to shared computation of joins and to cases in which the particular CJoin and DataPath join methods show good performance. Again, this constrained applicability of the CJoin and DataPath techniques has limited these systems to process only OLAP workloads so far, whereas SharedDB can be applied to transactional (i.e., OLTP), OLAP, and mixed workloads such as those modeled in the TPC-W benchmark. Furthermore, SharedDB can give response time guarantees which QPipe and DataPath cannot make. CJoin can only give response time guarantees under certain circumstances; i.e., if the bulk of the work is carried out as part of the Star join pipeline and not as part of other joins or as part of grouping, sorting, or aggregation. Section 3 discusses the commonalities and differences between SharedDB and QPipe, CJoin, and DataPath in more detail when the main ideas of SharedDB are described. In addition, Section 3.5 contains a detailed discussion of the advantages and disadvantages of SharedDB as compared to these related systems.

A specific problem of shared computation is that it may result in *deadlocks* in a pull-oriented query processor [6]. This problem can be alleviated by a *push-oriented* query processing approach which is the approach adopted by SharedDB. This push-based processing model has its roots in data stream processing technology; e.g., YFilter [8]. Even though it is applied to a totally different query processing paradigm, SharedDB has a number of additional commonalities with the YFilter approach: Just as YFilter, SharedDB compiles a set of queries into a single query plan and continuously uses the same plan for different generations of queries. In most applications, this approach becomes possible because the *kinds* of queries are known in advance and running an application over time involves executing the same queries with different parameter settings. The implementation of the TPC-W benchmark, for instance, involves about thirty different JDBC PreparedStatements that are executed with different parameter settings. The idea of such an *always-on-query plan* for short-lived queries was also adopted in the CJoin work [3] and is one of the commonalities between SharedDB, CJoin, and DataPath.

A push-based, dataflow architecture for query processing has also been adopted in the Eddies project [2]. While SharedDB and Eddies are also similar in a number of other technical details, the two systems were designed for totally different purposes. Eddies were designed for run-time adaptivity in situations in which the query optimizer cannot make good decisions at compile-time. Eddies, however, cannot provide any response time guarantees. In contrast, SharedDB is static and does not adapt the query plan at runtime. As a result, SharedDB will not always achieve the best possible response time for any given query, but SharedDB is able to provide response time guarantees which is the primary goal of this work.

Of course, there is a great deal of other related work on optimizing databases to meet SLAs in high-throughput situations. Examples include indexing, materialized views, and the design of advisors to get the best physical database design for a given workload [5]. Other examples include caching (e.g., [7]), reuse of query results (e.g., [18]), or optimization techniques that control the data placement in a distributed system (e.g., [20]). Various aspects of shared scans, one particular building block exploited in SharedDB (and QPipe, CJoin, and DataPath), have been studied in [10, 29, 21, 28]. Finally, the idea of bounded computation for constant-time query processing was pioneered in the Blink system [22].

## 3. SYSTEM OVERVIEW

This section presents the key ideas that combined provide the unique performance and predictability characteristics of SharedDB

**Result Set with Redundancy (First Normal Form)**

| Row_Id | Name | Other Attr. | query_id |
|---|---|---|---|
| 143 | John Smith | ... | 1 |
| 148 | Kate Johnson | ... | 3 |
| 143 | John Smith | ... | 2 |
| 148 | Kate Johnson | ... | 2 |
| 143 | John Smith | ... | 3 |

**Compact Result Set (NF2)**

| Row_Id | Name | Other Attr. | query_id |
|---|---|---|---|
| 143 | John Smith | ... | 1, 2, 3 |
| 148 | Kate Johnson | ... | 2, 3 |

**Figure 1: Query-Data Model Variants**

for a large range of different workloads: the data-query model, the global query plan, batched query execution, and shared operators. Furthermore, this section contains a detailed comparison of SharedDB versus its closest competitors. Implementation details of SharedDB are described in Section 4.

### 3.1 Data-Query Model

SharedDB features a novel data-query model to represent (shared) intermediary query results. This data-query model extends the relational data model by adding an additional column which keeps track of the identifiers of queries that are potentially interested in a tuple. Specifically, the schema of an (intermediary) Relation $R$ is represented as follows in this data-query model:

$$\{R_a, R_b, ..., R_n, \texttt{query\_id}\}$$

Here, $R_i$ are the (normal) attributes of $R$. The query_id attribute uniquely identifies each query that is currently active in SharedDB. As will be shown in the following sub-sections, the query_id can be used in relational operators just as any other attribute; for instance, it could be part of the join predicate between two relations (Section 3.3).

In a traditional, "first normal form" relational database system, the implementation of the data-query model would result in a great deal of redundancy. If a tuple is relevant for multiple queries, multiple copies of this tuple would have to be generated and processed; one for each relevant query. As a result, the complexity of query processing and the main memory footprint would grow linearly with the number of queries. The goal of SharedDB is to do better. Therefore, SharedDB implements this column internally as a *set-valued attribute* (i.e., using the NF2 model). As a result, an operator (e.g., a predicate) must be applied to a tuple only once independent of the number of concurrent queries and updates that may have subscribed to that tuple. Furthermore, the main memory footprint is reduced significantly. For illustration, Figure 1 shows examples of both formats.

There is a question of how to implement set-valued attributes most efficiently. In the literature, two data structures have been proposed: (a) bitmaps and (b) lists. For SharedDB, we chose to use a list-based implementation because that turned out to be the more space and time efficient option in all our experiments.

### 3.2 Global Query Plan

Instead of compiling every query into a separate query plan, SharedDB compiles the whole workload of the system into a single global query plan. The global query plan may serve hundreds or thousands of concurrent SQL queries and updates and may be reused over a long period of time, possibly for the entire lifetime of the system. As stated in Section 2, this approach was pioneered in the context of continuous query processing in data stream processing systems (e.g., YFilter [8]) and first applied to traditional, short-lived queries as part of the CJoin system [3].

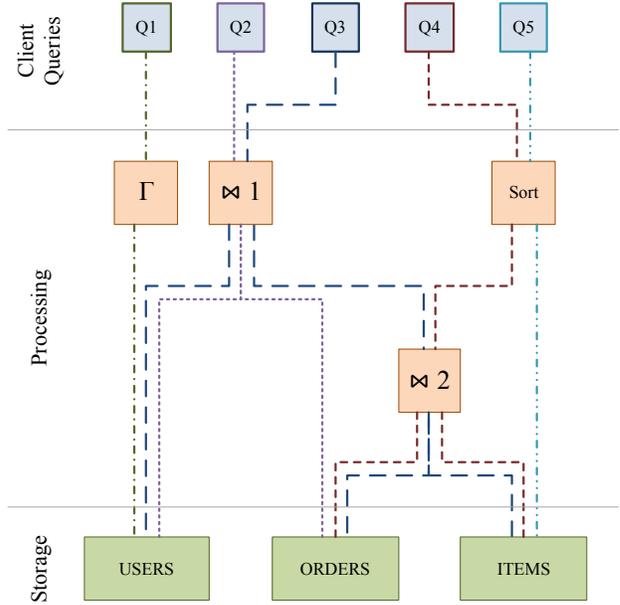

| Query | SQL |
|---|---|
| Q1 | SELECT COUNTRY, SUM(USER_ID) FROM USERS GROUP BY COUNTRY |
| Q2 | SELECT * FROM USERS U, ORDERS O WHERE U.USER_ID = O.USER_ID AND U.USERNAME = ? AND O.STATUS = 'OK' |
| Q3 | SELECT * FROM USERS U, ORDERS O, ITEMS I WHERE U.USER_ID = O.USER_ID AND O.ITEM_ID = I.ITEM_ID AND I.AVAILABLE < ? |
| Q4 | SELECT * FROM ORDERS O, ITEMS I WHERE O.ITEM_ID = I.ITEM_ID AND O.DATE > ? ORDER BY I.PRICE |
| Q5 | SELECT * FROM ITEMS I WHERE I.CATEGORY = ? ORDER BY I.PRICE |

**Figure 2: Example of a Global Query Plan**

Figure 2 shows an example of such a global query plan. In this example, four database operators are executed on three tables to evaluate five different query types. Figure 2 shows how different kinds of queries can share operators in different ways. For instance, both $Q_2$ and $Q_3$ involve a join between the *Users* and *Orders* table; consequently, this join is shared between queries of these two types (denoted as ⋈1 in Figure 2). Likewise, queries of types $Q_3$ and $Q_4$ share the join between the *Orders* and *Lineitems* table. The sort on *price* can be shared between queries of types $Q_4$ and $Q_5$.

Just as important as sharing across different types of queries is sharing within the *same* type of query. For instance, the plan shown in Figure 2 could be used to execute hundreds of concurrent queries of type $Q_4$ (in addition to hundreds of concurrent queries of the

528

other types), all with different parameter settings for the *O.DATE* predicate. All these concurrent $Q_4$ queries would share the same join and sort operators so that only a single big join and sort would be carried out for all these $Q_4$ queries.

To understand the execution model of SharedDB, it is important to define what *concurrent* means. As mentioned in Section 2, SharedDB facilitates the sharing of operators in a very different way than QPipe, CJoin, DataPath, and other related systems that support operator-level sharing of computation. These systems start executing each query as soon as it arrives. Unfortunately, this approach limits the opportunities to share computation for many database operators as observed in [15]. In order to overcome these limitations, CJoin and DataPath devise specific join methods. In contrast, SharedDB *batches* queries and updates; it is this batching that enables SharedDB to exploit shared computation in a scalable and generic way, thereby making use of traditional, best-of-breed algorithms to implement joins, sorting, and grouping.

SharedDB batches queries and updates in the following way: While one batch of queries and updates is processed, newly arriving queries and updates are queued. When the current batch of queries and updates has been processed, then the queues are emptied in order to form the next batch of queries and updates. Metaphorically, SharedDB works like the blood circulation: With every "heartbeat", tuples are pushed through the global query plan in order to process the next generation of queries and updates. For OLTP workloads, these heartbeats can be frequent in the order of one second or even less.

The queueing of queries is fine-grained per input relation and query operator: In the example of Figure 2, queries of Type $Q_1, Q_2$, and $Q_3$ would queue for reading the *Users* relation, queries of Type $Q_2, Q_3$, and $Q_4$ would queue for reading the *Order* relation and so on. Then *User* tuples (generated for all three query types) would queue for the $\Gamma$ and $\bowtie 1$ operators which in turn would process these tuples in batches so that all *User* tuples belonging to a specific query are processed within a single batch. Pipelines are only created for certain operators; for instance, an operator can stream its output into the *build* phase of a hash join and even then, tuples are processed in batches following a *vector model* of execution for better instruction cache locality [14, 17] (Section 4).

SharedDB works particularly well if most query types are known in advance; e.g., as part of JDBC PreparedStatements. In this case, SharedDB can generate a global query plan for all the query types known in advance and exploit sharing in the best possible way for these queries. Ad-hoc queries need to be processed individually. Nevertheless, even ad-hoc queries can take advantage of sharing. For instance, an ad-hoc query that asks for the ten users that have placed the most orders could share $\bowtie 1$ with all queries of type $Q_2$ and $Q_3$ in the global plan of Figure 2. The Top 10 operation of that ad-hoc query, however, would have to be compiled and executed separately just as in any other traditional database system. In some sense, all operators of the global plan can be regarded by the query compiler as materialized views which are available to speed-up the processing of ad-hoc queries. This observation has also been made in the QPipe project [15].

### 3.3 Shared Join Plans

One of the key innovations of SharedDB is the way it processes joins, sorts, and group-bys. Figure 3 shows how SharedDB processes joins for three different queries ($Q_1, Q_2, Q_3$). All three queries involve a join between tables $R$ and $S$, but each query has separate predicates on the $S$ and $R$ tables. In the first step, each query is parsed and compiled individually, thereby pushing down predicates. This step is typically referred to as *logical query op-*

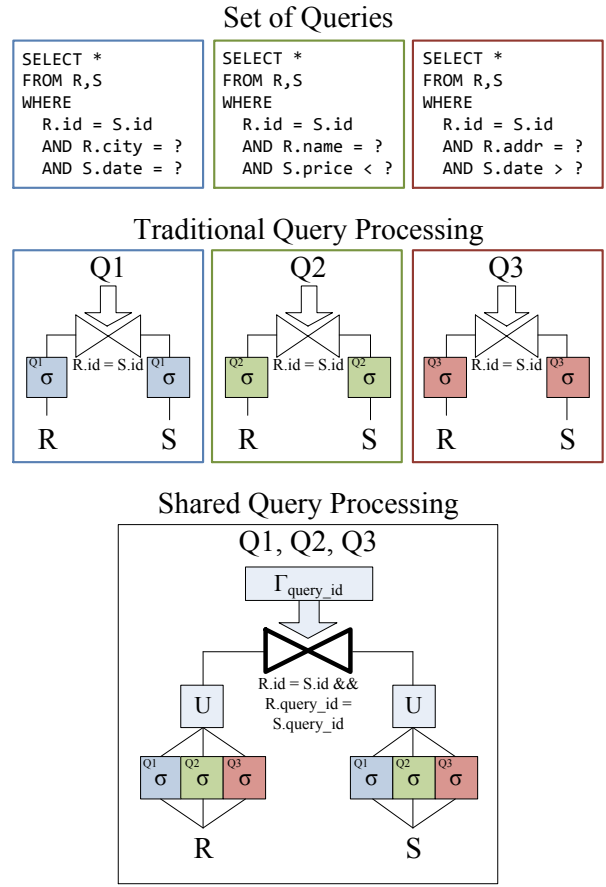

**Figure 3: Shared Join**

*timization* [13]. The result of this logical optimization, one plan for each query, is shown in the middle of Figure 3. In the second step, the three individual query plans are merged into a single global plan. That is, rather than processing three *small* joins (one for each query), one *big* join is executed that meets the requirements of all three queries. Technically, the *union* of all $R$ and $S$ tuples that the three queries are interested in are considered as part of the join. Furthermore, the join predicate is amended, thereby considering the query_id. This way, an $R$ tuple that is only relevant for Query $Q_1$ does not match an $S$ tuple that is only relevant for Query $Q_2$. Finally, the routing of the join results to the relevant queries is carried out using a *grouping operator* ($\Gamma$) by query_id.

As stated in the introduction, this way to process joins sounds like a bad idea at first glance: It is usually better to process a few small joins than to process one big join. This approach only becomes advantageous if there are many (possibly hundreds) of concurrent queries and there is overlap in the tuples that need to be processed for a set of queries. One particularly nice way in which the global join plan of Figure 3 supports scalability with the number of concurrent queries is by making the query_id part of the join predicate. This way, the query_id can be indexed; for instance, a hash join could be used that builds a hash table on the query_id of $S$ (see below). This observation shows nicely how SharedDB achieves scalability by turning queries into data that can be processed using traditional query processing techniques. This is another key idea that SharedDB has adopted from data stream processing systems.

Another crucial advantage of the global join plan of Figure 3 is that any join method can be used; e.g., hashing, sorting, index-

529

based, and nested-loops. In particular, any parallel or cache-aware join methods can be used (e.g., [19]). Furthermore, as observed in the previous paragraph, either $R.id = S.id$ or $R.query\_id = S.query\_id$ can be used as primary join predicates. If the latter, a set-based join is carried out as studied in [16]. In our implementation, we use a simple hash table that maps a `query_id` to a set of pointers that reference the corresponding tuples. Again, this set of pointers is implemented using a *list* data structure because this particular join method is only beneficial if these sets are small. It is even possible to use multi-dimensional join methods which are commonly used in spatial and temporal database systems; e.g., [9]. In summary, SharedDB can always make use of the best-of-breed algorithms. In contrast, the processing model of CJoin and DataPath constrains the use of join methods used: In some cases, using the dedicated join method of DataPath might show good performance; in other cases, however, its performance may be terrible. SharedDB makes use of multiple join methods for the same reasons as traditional database systems.

In addition to supporting multiple join methods, SharedDB does not constrain join ordering thanks to its flexible data-query model. In Figure 2, for instance, the *Orders / Items* join ($\bowtie 2$) of $Q_3$ could be carried out before or after the join with *Users* ($\bowtie 1$). As a result, the following join odering for $Q_3$ would also be possible:

$$(Users \bowtie Orders) \bowtie Items$$

This join order for $Q_3$ would also enable sharing $\bowtie 1$ with $Q_2$ and $\bowtie 2$ with $Q_4$, in the same way as the original join order for $Q_3$ depicted in Figure 2. To share a join across queries, SharedDB only fixes the join method and the inner and outer relation of the join for all queries that are share the join.

Studying the details of the SharedDB query compiler and optimizer is beyond the scope of this paper. Developing a sophisticated cost-based optimizer is an important avenue for future work. We believe that the two-step optimization approach shown in Figure 3 (i.e., determine join orders for each query individually and then merge the plans for the individual queries into a single global plan) will perform well for many workloads. However, it is conceivable that an approach that optimizes the global plan for all queries in a single pass results in better performance in certain cases.

### 3.4 Other Operators: Scan, Sort, Group-by

The idea shown in Figure 3 to process shared joins can also be applied to process any other operator of the relational algebra. Figure 4 illustrates the principle for a shared sort using two example queries and a few tuples of a *Users* table. Again, in theory, it is better to have a few small sorts than one big sort, but sharing may more than offset this effect. In this example, it is more efficient to do *one* sort with four tuples than to do *two* sorts with three tuples each. Obviously, the overlap increases with the number of queries.

The Top-N operator is an extension of the sort operator and can, thus, benefit from sharing in a similar way as the sort operator. In the SharedDB implementation, the shared Top-N operator first sorts all the tuples that are relevant for all the active queries; thus, the sorting is shared. Then, it filters the Top N results for each query individually.

Like Top N, the Group-By operator is carried out in two phases. In the first phase, the input tuples are grouped. Again, this phase can be shared so that all the tuples that are relevant for all active queries are grouped in one big batch. Also, any kind of grouping algorithm (e.g., hashing or sorting) can be used for this purpose. In the second phase, `HAVING` predicates and aggregation functions are applied to the tuples of each group. Just as for Top N, this second phase must be carried out for each query individually. Fortunately,

**Shared Sort Queries**

| Query A: | `SELECT * FROM USERS`<br>`WHERE BIRTHDATE > 1980.01.01`<br>`ORDER BY NAME` |
|---|---|
| Query B: | `SELECT * FROM USERS`<br>`WHERE ACCOUNT > 1000`<br>`ORDER BY NAME` |

**Relation: USERS**

| Name | Account | Birthdate | Query_Ids |
|---|---|---|---|
| John Smith | 3,000 | 1980.03.05 | A, B |
| Kate Johnson | 800 | 1976.04.11 | |
| Bill Harisson | 1,230 | 1978.03.02 | B |
| Nick Lee | 540 | 1982.02.09 | A |
| James Meyer | 2,300 | 1981.03.09 | A, B |

**Sorted Output**

| Name | Account | Birthdate | Query_Ids |
|---|---|---|---|
| Bill Harisson | 1,230 | 1978.03.02 | B |
| John Smith | 3,000 | 1980.03.05 | A, B |
| James Meyer | 2,300 | 1981.03.09 | A, B |
| Nick Lee | 540 | 1982.02.09 | A |

**Figure 4: Shared Sort**

grouping is the most expensive part so that sharing can be applied to reduce the cost of the most performance critical operation.

The storage manager of SharedDB provides two operators for accessing tables: *shared table scans* and *shared index probes*. We do not claim any novelty here. Shared table scans have been studied extensively in previous work ([10, 29, 21, 28]) and SharedDB uses the ClockScan algorithm proposed in [28]. Shared index probes have also been studied in the past (e.g., [12]) and SharedDB simply adopts the well established techniques in that domain, too. Both operators generate tuples in the data-query model such as those shown in Figure 1.

### 3.5 Discussion

This section summarizes the main advantages and disadvantages of SharedDB in comparison to traditional, query-at-a-time systems and other systems that exploit shared computation.

*SharedDB vs. "query-at-a-time"*. The main *dis*advantage of SharedDB is that it adds latency to each query due to its batch-based execution model. In contrast, traditional database system start processing queries as soon as they are submitted by an application. In the worst case, batching increases latency by a factor of 2: one cycle of queuing and one cycle of actual query processing.

The biggest *advantage* of SharedDB is that it is able to bound computation and scales with the number of concurrent queries and updates. In the worst case if there are many concurrent queries that involve all the tuples of the whole database, SharedDB joins and sorts the *whole* relations as part of the global query plan, independent of the number of concurrent queries. This way, SharedDB can give response time guarantees which is critical for many modern applications to meet SLAs. For instance, if SLAs specify that all queries must be processed within 3 seconds, then SharedDB would provision enough CPU cores such that a batch of queries can be processed in at most 1.5 seconds in the worst case. All this is predictable and can be planned upfront: There is no interference and resource contention between concurrent queries because SharedDB schedules the data flow and the utilization of cores at compile-time as part of its global plan. In contrast, the work carried out by traditional database systems grows linearly with the number of concurrent queries. Furthermore, traditional database systems allocate a separate thread for each query and these threads might compete for shared resources (e.g., the main memory bus or processor caches) in an unpredictable and uncontrollable way.

As mentioned in Section 3.3, SharedDB might result in extra work if the load on the system is light and there is little or no overlap in the data processed by the queries that share a common operator. With an increasing load, however, the overlap increases. In theory, if each Query $Q_i$ needs to process $n_i$ tuples, there are $k$ concurrent queries, $n = \Sigma_{i=1}^{k} n_i$, and $o$ is the number of tuples that at least one query needs to process ($o \leq n$), then SharedDB will save work for an operator with complexity $\mathcal{O}(f(n))$ if:

$$f(o) < \Sigma_{i=1} k f(n_i)$$

For operators with *linear* complexity (e.g., table scans and joins under certain circumstances), this equation is always fulfilled (unless $o = n$ which is the worst case for SharedDB). For operators with a complexity of $f(n) = n * log n$ (e.g., sorts and certain joins), the advantage of SharedDB depends on $o$ and $n$: In such cases, SharedDB will result in extra work in the worst case ($o = n$). But, even in these bad cases, SharedDB maintains its property of bounded computation and predictable performance.

*SharedDB vs. "pipelined sharing"*. The most significant *dis*advantage of SharedDB as compared to other, recent approaches to effect shared computation (e.g., QPipe, CJoin, and DataPath) is again that SharedDB adds latency due its batched processing model. In contrast, QPipe, CJoin, and DataPath have a continuous query processing model. The big advantage of this batched processing model in concert with the special way in which joins and other operators are processed, is its generality to any kind of operator of the relational algebra, any kind of algorithm (e.g., join method), and to the processing of updates. This generality enables SharedDB to process any kind of workload (OLTP, OLAP, and mixed) without any special tuning, whereas QPipe, CJoin, and DataPath have so far only been shown to work well for OLAP workloads with queries that each involve processing a large portion of the entire database. As observed in the QPipe work [15], the kinds of sharing are limited depending on the query operator in a continuous query processing model. As a result, CJoin and DataPath rely on specific, dedicated join methods in order to carry out shared join computation; other operators (e.g., sorting and grouping) need to be carried out for each query individually in these systems. Furthermore, these dedicated join methods only show good performance for specific kinds of workloads. Just as a traditional database system, SharedDB supports multiple join methods in order to adapt to different kinds of workloads.

A specific advantage of SharedDB as compared to QPipe and DataPath is its ability to meet SLAs and bound the response time of queries. Predictable performance is also supported in CJoin, but only for star-join queries. Another advantage of the batched processing model of SharedDB is that it supports concurrent updates and strong consistency (e.g., Snapshot Isolation). Section 4 gives more details on how updates are processed in SharedDB. Again, updates and transaction processing have not been studied in the context of QPipe and DataPath; CJoin does support concurrent updates, but the model is more complicated than in SharedDB.

## 4. IMPLEMENTATION DETAILS

This section describes a number of implementation details of SharedDB that will help the reader get a better understanding of the system.

### 4.1 Query Model

As described in Section 3, SharedDB evaluates queries using a data flow network of always-on database operators. There are no individual query plans and instead a global plan is always active.

Every SharedDB query describes an acyclic path in the data flow network. As an example, Figure 5 shows a possible representation of Q1 of Figure 2. In this example, the GroupBy operator receives a query from the Output operator. In order to execute it, another query is issued to the TableScan operator. Result tuples are generated by the TableScan, passed to the GroupBy operator and finally sent to the clients.

|    | Operator  | Configuration                          |
|----|-----------|----------------------------------------|
| 1. | TableScan | USERS WHERE LAST_LOGIN > 2011.01.01    |
| 2. | GroupBy   | GROUP(USERS.COUNTRY)                   |
|    |           | SUM(USERS.ACCOUNT)                     |
| 3. | Output    | Network, TCP Port 5843                 |

**Figure 5: An Example of a Query in SharedDB**

### 4.2 Operators

Shared operators are designed to evaluate a number of queries concurrently by processing them in *cycles*. Every cycle evaluates the set of active queries or subqueries. Any additional queries that arrive after the cycle has started, are queued and will be evaluated during the next cycle.

Algorithm 1 shows an abstract SharedDB operator that processes queries in cycles. At the beginning of the cycle, the operator dequeues the pending queries and activates them by issuing their subqueries to the respective operators. Then it receives the generated result tuples they generated, processes them and forwards the processed output to the issuers of the queries. For example, in the case of a filter operator, like SQL LIKE, the ProcessTuple function tests the tuple against the LIKE expression. If it matches, the result tuple is pushed to the next operator in the pipeline.

Blocking operators, such as the SORT operator, can use the function ProcessTuple to append the tuple to a buffer structure (i.e., a vector). The same buffer structure is used for all the queries that belong to the same batch. In this case, no results are produced from ProcessTuple. Once all the result tuples have been received, the buffer structure is sorted and it is pushed to the consumers as part of the SendEndOfStream function.

### 4.3 Runtime Configuration

All database operators are executed in a separate hardware context. If there are enough CPU cores in the system, each database operator is assigned to a different CPU core, using hard processor affinity. This guarantees that the threads do not migrate between processors, allowing for optimal instruction cache locality. Additionally, the binding of operators to physical CPU cores allows for optimal use of NUMA (non-uniform memory access) hardware architectures. The local stack and heap of every operator is allocated on memory that has the minimum NUMA distance. As a result CPUs never access each other's local memory except for passing operation results, giving maximum bandwidth and minimum access latency where it matters: during query processing.

**Algorithm 1**: Skeleton of a SharedDB Operator

```
Data: SyncedQueue iqq;            // incoming queries queue
Data: SyncedQueue irq;            // incoming result tuples queue
while true do
    Array aq ← ∅ ;                           //Array of active queries
    //Operators that will receive the produced results
    Array consumers ← ∅ ;
    //Activate all queries in the incoming queue
    while ¬IsEmpty(iqq) do Put(aq, Get(iqq));
    //Enqueue subqueries to underlying operators
    foreach Query q ∈ ag do
        Query subQuery ← GetSubQuery(q);
        Operator op ← GetOperator(subQuery);
        EnqueueQuery(op, subQuery);
        Operator consumer ← GetConsumer(q);
        Put(consumers, consumer);
    //Loop until all active queries have finished
    Number queriesLeft ← Size(aq);
    while queriesLeft do
        //Receive a tuple from the underlying operators
        Tuple t ← Get(irq);
        //Process the incoming tuple. Depending on the
           operator, this function might generate results
        Tuple resultTuple ← ProcessTuple(t);
        if ¬IsNull(resultTuple) then
            SendResult(consumers, resultTuple);
        if IsEndOfStream(t) then
            queriesLeft ← queriesLeft − 1;
    //Notify the consumers that processing has finished
    SendEndOfStream(consumers);
```

Moreover, careful deployment of database operators across CPU cores can further exploit modern multi core hardware architectures. In these systems, CPU cores that are located on the same chip, share components of the memory architecture, like the L3 cache and the NUMA main-memory region. Database operators that access similar sets of records can be assigned to "adjacent" CPU cores in order to benefit from the sharing of these memory components.

Our existing implementation supports all these optimizations. Currently the assignment of operators to CPU cores has been performed manually by examining the data access paths of every operator. Future versions of SharedDB will support an optimizer to automatically deploy operators to the proper CPU cores.

### 4.4 Storage Manager and Transactions

The current implementation of SharedDB is based on the Crescando storage manager which has been successfully deployed to serve demanding update-intensive operational business intelligence workloads of the travel industry [28]. The Crescando storage manager is currently in production at Amadeus, the market leader for airline reservation.

In the version used for SharedDB, Crescando supports two access methods for reading base tables: (a) (shared) table scan, and (b) index probes. Table scans are carried out using the ClockScan algorithm, described in detail in [28]. ClockScan batches queries and updates in the same way as SharedDB and processes a whole batch of queries and updates. As a result, the ClockScan algorithm fits nicely into the overall SharedDB design. Performance is increased by indexing the query predicates instead of the data and performing query-data joins, a technique widely used in data stream processing. Updates are executed in arrival order as part of the same scan that executes the queries. At this level, Crescando guarantees that all select queries read a consistent snapshot of data.

The original Crescando storage manager presented in [28] only supports full table scans using the ClockScan algorithm. For this work, we extended Crescando and implemented B-Tree indexes and index probe operators as an additional access path. These index probe operators are used to implement regular scans (with predicates) on base tables and to implement index nested-loops joins. The logic of the index probe operator is similar to Algorithm 1. Look-ups are enqueued in the pending query queue which is emptied at the beginning of each cycle. During the cycle, the updates are executed in the arrival order and multiple B-Tree look-ups are used to evaluate all the select queries. Executing multiple look-ups in one cycle allows for better instruction and data cache locality [12]. Just as the (shared) full table scan, the index probe operator guarantees that all select queries will read a consistent snapshot.

In general, transactions can be implemented in SharedDB in almost the same way as in any other database system. In particular, atomicity, consistency (i.e., checking integrity constraints), and durability (i.e., recovery) are completely orthogonal to shared query and update processing. With regard to isolation, the design of SharedDB favors optimistic and multi-version concurrency control because any kind of locking would result in unpredictable response times due to lock contention and blocking. In particular, Snapshot Isolation, as supported by the Crescando storage manager, complements nicely the batch-oriented shared query processing model of SharedDB. With regard to durability, Crescando keeps all data in main memory, but it also supports full recovery by checkpointing and logging all data to disk.

Crescando supports horizontal partitioning of data and processing several partitions with different cores in parallel. This feature of Crescando, however, was not used in the performance experiments presented in Section 5.

### 4.5 Replication

The design of our system allows replication in SharedDB. In fact, not only storage operators but also database operators can be replicated. Replicating a storage operator does not hurt consistency, because updates are always executed in the same order as they were received by the system. Moreover, data processing operators can be also replicated in order to reduce the effects of bottlenecks and hotspots in the data flow network. For instance, if a specific operator becomes a bottleneck, SharedDB can partition the load across two replicas of the same physical operators. Similar to the deployment of database operators to CPU cores, replicating a specific operator is a task that needs to be performed by an optimizer, based on the global query plan.

## 5. PERFORMANCE EXPERIMENTS AND RESULTS

In order to evaluate the performance of SharedDB, we carried out a series of performance experiments. As baselines, we used MySQL and a high-end commercial database product. We present results of comprehensive workloads using the TPC-W benchmark and the results of micro benchmarks with individual queries in order to demonstrate specific effects.

### 5.1 Experimental Environment

The TPC-W benchmark models an online bookstore. It assesses the performance of multi-tier information systems which contain a database layer, a web server layer and a client layer. Every client of the system is an emulated browser that issues http requests (web interactions in TPC-W terminology) to the web server layer. Between two web interactions, there is a "thinktime" which is defined by a negative exponential distribution and has an average of 7 seconds. The web servers accept the clients' requests and issue queries to the database layer in order to retrieve the requested data. Each client interaction is translated to a number of database queries, depending on the type of the interaction. For instance, to execute the "Home" web interaction (i.e., a user visits the application's home page), two queries have to be evaluated: The first query fetches a set of promotion items, and the second query retrieves the profile of the user.

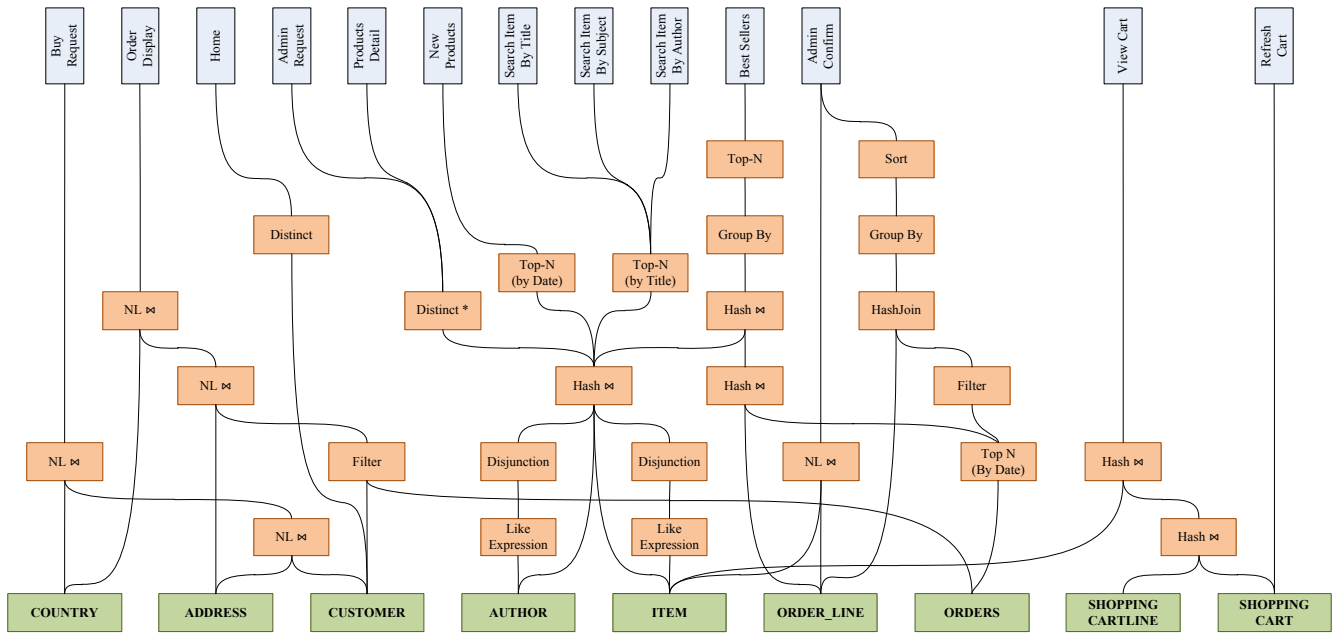

Figure 6: Global Plan for the TPC-W Benchmark (Updates have been omitted for simplicity)

In order to be compliant with the TPC-W specification, every web interaction needs to be answered in a predefined amount of time. The timeout depends on the type of the web interaction, ranging from 2 seconds for small, point queries, up to 20 seconds for long running analytical queries. Any web interaction that exceeds this timeout, is not valid.

TPC-W contains a total of 14 different web interactions. Every web interaction has a different probability of appearing in the workload. The probabilities of all the web interactions are given by the "workload mix". TPC-W defines three different workload mixes: Browsing, Shopping, and Ordering. The Browsing mix is a read-mostly, search intensive workload with few updates and many analytical queries. The Ordering mix is a write-intensive workload with only a few analytical queries. The Shopping mix is somewhere in between with some updates and some analytical queries.

In all experiments reported in this paper, we used a 48 core machine as a database server. This machine features four twelve-core AMD Opteron 6174 ("Magny-Cours") sockets and is equipped with 128 GB of DDR3 1333 RAM. Each core has a 2.2 GHz clock frequency, 128 KB L1 cache, 512 KB L2 cache, and is connected to a shared 12 MB L3 cache. The operating system used in all experiments was a 64-bit SMP Linux. The emulated browsers were executed on up to eight client machines, each having 16 CPU cores and 24 GB of DDR3 1066 RAM. The clients also ran the application logic; that is, the clients issued queries directly to the database server. This is a slight simplification of the TPC-W set-up and justi-fied because we were interested in the performance of the database system under high load. The client machines were connected to the database server machine using a 1 Gbps ethernet.

We implemented the full TPC-W workload in SharedDB. Figure 6 shows the global query plan that was generated. It consists of 26 database operators in addition to shared scans and index probe operators to access the nine base tables of the TPC-W benchmark. The current version of SharedDB does not feature any parallel join or sort algorithms and does not support partitioning or replication of base data. As a result, we used at most 32 cores for SharedDB, one core for each operator and nine cores for the shared table scans. (Filter operators do not require a separate CPU core and the `Distinct *` operator is evaluated as part of the underlying `Hash⋈` operator.) SharedDB is able to make use of the additional CPU cores by replicating operators, as explained in Section 4.5. However, we avoided using replication of data and operators in all our experiments as the goal is to study the benefits of sharing rather than the benefits of this particular technique. In order to demonstrate the scalability of SharedDB, we varied the number of cores between 1 and 32 in some experiments. We used the kernel parameter `maxcpus` to limit the number of available CPU cores. If not stated otherwise, 24 cores were used.

In all experiments reported in this paper, SharedDB held all the data in main memory. Disk I/O was only required to log updates as part of the Crescando storage manager (Section 4.4). As a result, running the TPC-W benchmark on SharedDB was CPU-bound. It was also CPU-bound for the two baseline systems that we describe in the next sub-section.

### 5.2 Baselines

To put the performance of SharedDB into perspective, we compared it against two existing database systems. The first one is a popular commercial system that will be referred to as SystemX. The second one is a widely used database system, MySQL 5.1 using the InnoDB storage engine. Just like SharedDB, MySQL and SystemX are general-purpose database systems, designed to perform well for any kind of workload. Other existing solutions may reach better performance for specific workloads; e.g., column stores for OLAP workloads and lock-free, single-threaded systems for OLTP workloads. We do not compare SharedDB to these systems because we wanted to understand specifically the performance tradeoffs of SharedDB's shared execution model as opposed to the traditional, query-at-a-time model.

Clearly, the comparison of SharedDB with MySQL and SystemX is not an apples-to-apples comparison, but we tried to make the comparison as fair as possible by fine-tuning the two baseline systems. We built all the necessary indexes on the two systems and we used an in-main-memory filesystem for the data files of both systems. Furthermore, we provided a big memory buffer pool for MySQL and SystemX, big enough to hold the database and all indexes. Finally, we filled this buffer pool by carrying out full table scans and a warm-up phase of the benchmark. As a result, neither MySQL nor SystemX performed any disk I/O to carry out queries. Disk I/O was only performed by SystemX and SharedDB as part of logging in order to persist updates. For MySQL, all recovery options were turned off so that MySQL did not even perform disk I/O for logging. Overall, however, the workload was CPU-bound for all three systems. The isolation level used was "read committed", as TPC-W requires only session consistency.

We disabled query and result caching in both MySQL and SystemX. Query caching allows a database system to skip execution of a query as long as it has been executed before and no change in the dataset has occurred. Result caching performs the same optimization on each individual subquery. By disabling any caching, we guarantee that we measure the performance of evaluating every query, rather than fetching pre-calculated results from a cache.

In summary, we did everything to make a fair comparison between the three systems. Overall, however, our goal is not to show that SharedDB is better than MySQL, SystemX, or any other existing database system. Our goal was to show that batch-oriented sharing can result in predictable performance under high load. To this end, we also present the results of micro-benchmarks that study the behavior of all systems under growing load and isolates the effects of "query-at-a-time" vs. "shared" query processing.

Unfortunately, we could not carry out performance experiments on QPipe, CJoin, or DataPath, the closest competitors of SharedDB in terms of shared computation. These systems are research prototypes and were not available for experimentation. Furthermore, we believe that significant research is necessary before the special, continuous (as opposed to batch-oriented) sharing featured by these systems can be applied to comprehensive workloads such as the TPC-W benchmark. So far, these systems have only been applied to OLAP queries of the TPC-H and SSB benchmarks.

## 5.3 Performance under Varying Load

In the first set of experiments, we compared the performance of SharedDB, MySQL and SystemX on the three workload mixes of TPC-W. We varied the load of the system by increasing the number of emulated browsers and measured the web interactions that were successfully answered by the system in the response time limit that is defined by the TPC-W specification. All web interactions that exceeded this limit were not accounted as successful. For this experiment, we configured the database system to use 24 CPU cores.

Figure 7 shows the results. SharedDB is able to achieve higher throughput in all three TPC-W workload mixes. For instance, in the Browsing mix, we see that SharedDB is able to sustain twice the throughput of SystemX and eight times the throughput of MySQL. The Browsing mix involves use cases of customers searching for items. Most of the search queries are heavy queries that involve a number of joins and sorts. This result confirms previous results with MQO on TPC-H queries (e.g., [15]) that shared computation is beneficial if heavy queries need to be executed.

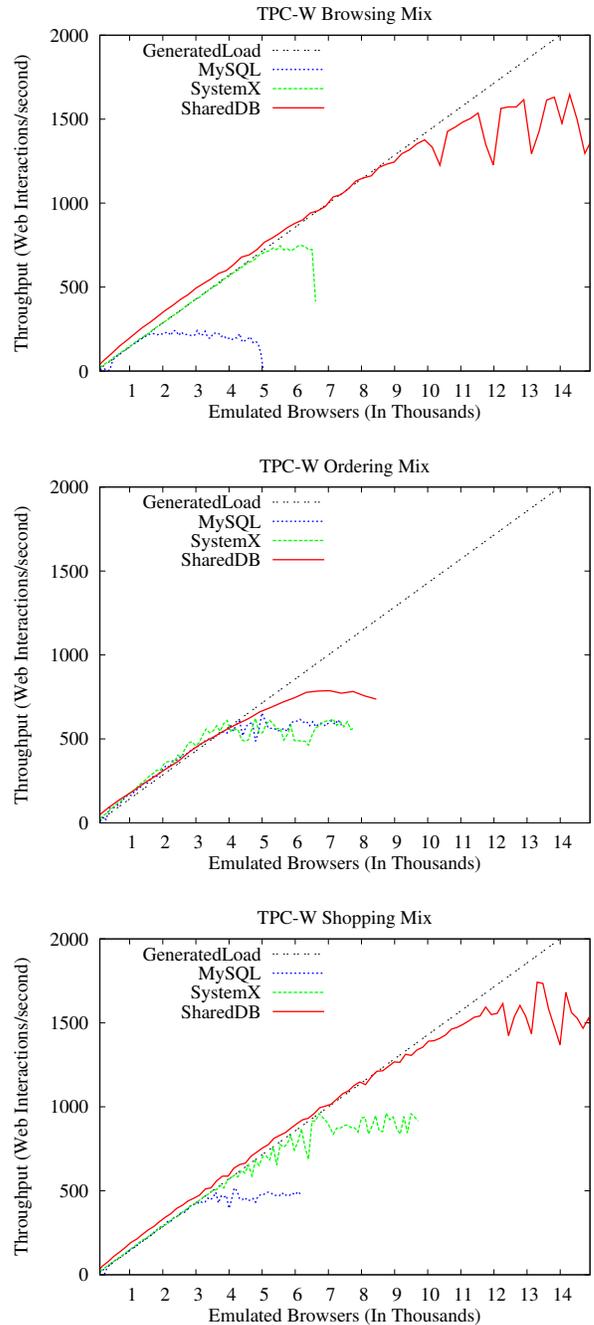

Figure 7: Throughput: Varying Load, All Mixes

The Ordering mix involves few searches for items, and instead places many new orders. For this mix, MySQL is able to execute as many web interactions as SystemX. Again, MySQL had a bit of an unfair advantage as compared to SystemX and SharedDB here because all recovery options were turned off for MySQL. SharedDB still wins in this experiment, but the margins are lower. In the Ordering mix, most queries are point queries that can be executed highly efficiently with an index look-up in a traditional, query-at-a-time fashion. Furthermore, there is a little benefit for sharing for such point queries.

Finally, the bottom diagram of Figure 7 shows the throughput results with varying load of the three systems for the Shopping mix. Again, SharedDB is the clear winner and again the reason is that

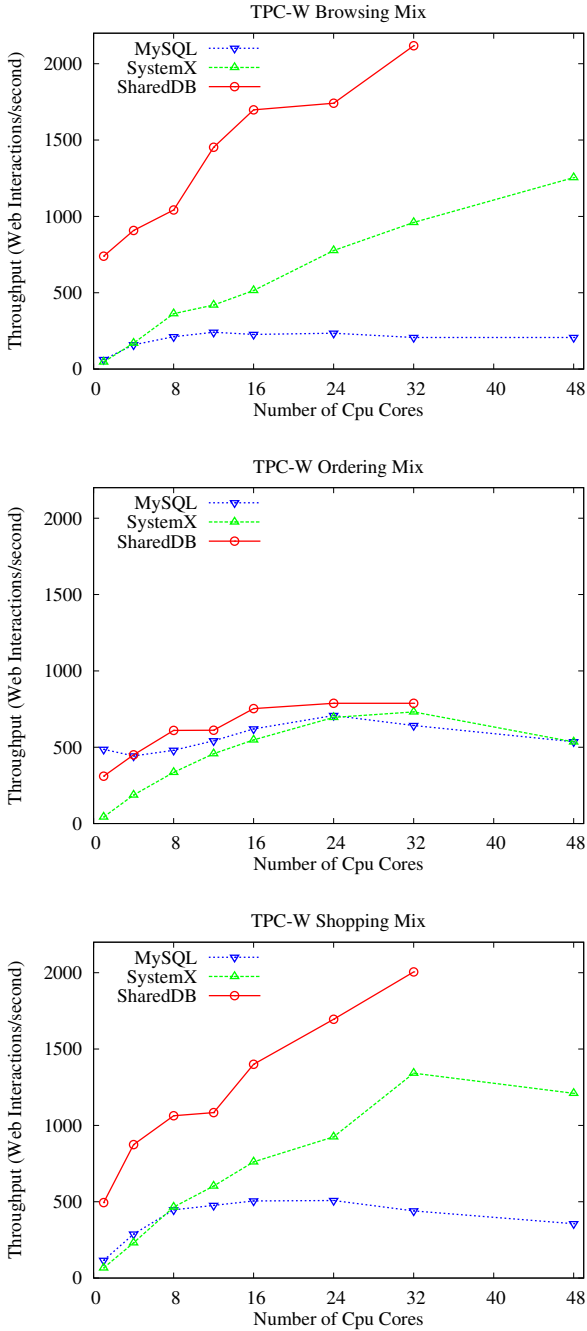

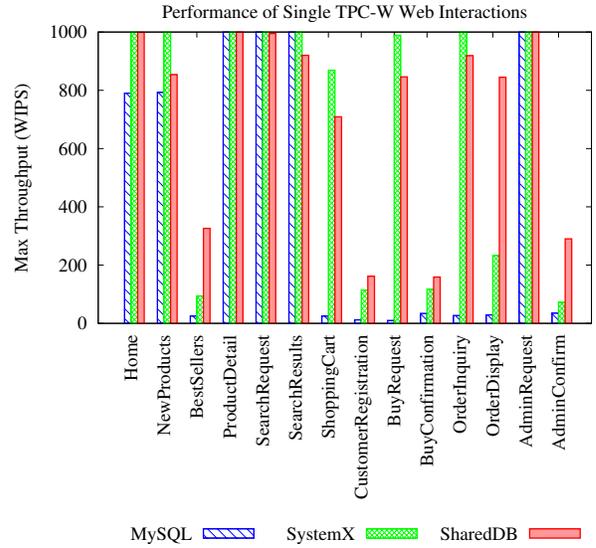

Figure 9: Analysis of Individual Web Interactions

### 5.4 Scaling with the Number of Cores

In the second set of experiments, we explored the impact of having additional CPU cores on the database server. For this reason, we varied the number of available CPU cores of the database server machine from 1 to 48 and repeated the experiments of Section 5.3 for each configuration. As mentioned in Section 5.1, we varied the number of cores for SharedDB only between 1 to 32 because SharedDB cannot take advantage of additional cores for the TPC-W benchmark in the configuration used for these experiments. We measured the maximum number of successful web interactions per second each system could achieve.

Figure 8 shows that SharedDB again is the clear winner for all three workload mixes and almost independent of the number of cores. SharedDB is only outperformed by MySQL for the Ordering mix if the database server is constrained to using only a single core. Again, the magic ingredient is sharing and the special architecture of SharedDB. Both SharedDB and SystemX scale nicely with the number of cores for the Browsing and Shopping mixes. Scalability is limited for the update-intensive Ordering mix because at some point concurrency control and transaction management limit the throughput of the system. MySQL does not scale beyond twelve cores, independent of the workload. This observation was also made in a recent study by Salomie et al. [23].

### 5.5 Analysis of Individual Web Interactions

The TPC-W benchmark involves a variety of different web interactions, each involving a different set of queries. For instance, the home web interaction involves two simple point queries (fetching promotion articles and a user's profile). Other web interactions involve point queries and several updates. Finally, there are also web interactions that involve heavy, analytical queries with multiple joins, grouping, and sorting. Figure 9 shows the maximum throughput that each of the three systems can achieve if the clients are configured to issue only queries that correspond to a single web interaction. These experiments were carried out in a configuration with 24 cores for the database server.

Figure 8: Max. Throughput: Vary # Cores, All Mixes

SharedDB takes advantage of shared computation for heavy and medium-sized queries.

We would like to reiterate that SharedDB used the *same* global plan for all three mixes. SharedDB was not adapted or tuned in any way to meet the specific requirements of these workloads. Obviously, a "query-at-a-time" system can adapt much better to the workload and, for instance, optimize a query based on the currently available resources. We do not know how, for example, SystemX does so. It is clear, however, that shared computation is more critical to sustain high throughputs with response time guarantees than any adaptation technique implemented in MySQL and SystemX.

SharedDB wins in this experiment for many kinds of web interactions (e.g., BestSellers and CustomerRegistration). Again, sharing is the main reason for SharedDB's success. Keep in mind that SharedDB does not only support sharing across queries of different types but also sharing between concurrent queries of the same type, executed with different parameter settings. Figure 9, however, also shows that SharedDB loses for several web interactions as compared to SystemX (e.g., NewProducts and ShoppingCart). These web interactions involve mostly point queries and/or updates for which sharing does not help much. SystemX wins because it is the more mature system and carries out the same work more efficiently than SharedDB. In other words, SharedDB can only beat SystemX if it carries out *less work* as a result of shared computation.

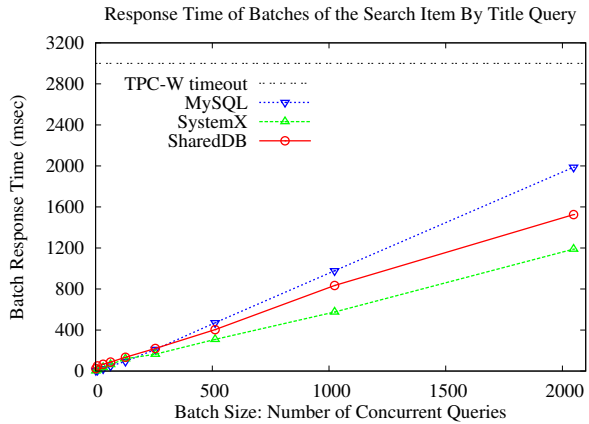

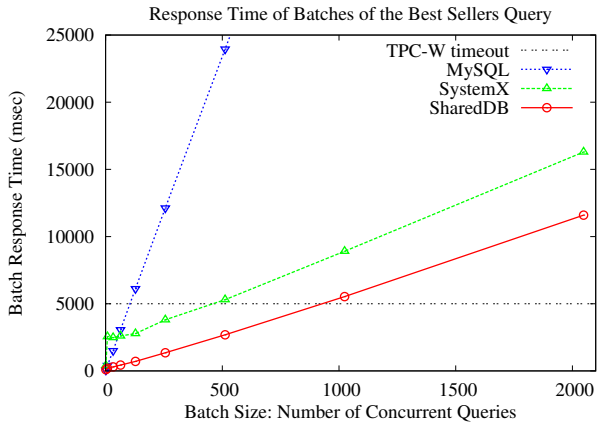

**Figure 10: Heavy Queries vs. Light Queries**

### 5.6 Heavy Queries vs. Light Queries

Next, we analyzed the performance of the three test systems under two very different queries of the TPC-W benchmark. The first one uses a key identifier to select an item and its author. This query is part of the `ProductDetail` web interaction. It is a lightweight query that performs a join of two relations and fetches one record from each of them. The second query is the "best sellers" query that is part of the `BestSellers` web interaction. This heavy query involves the analysis of the latest 3,333 orders that have been placed by customers in order to retrieve the most ordered items that match a selection predicate provided by the client. It performs three joins over four relations which are followed by a group-by operator and additional sorting of the results.

We used batches of an increasing number of such queries and issued a stream of them to the three systems while measuring the time needed in order to complete the whole batch. For SharedDB, the measured time includes the queueing time of the batch, as described in Section 3.2. The results are shown in Figure 10. With regard to the search item query, the performance of the three sys-tems follows the same trend. SystemX is able to execute the batches of queries faster than SharedDB, which is expected as explained in Section 5.5. This query is executed so fast that the overhead of batching queries and updates is greater than the gains.

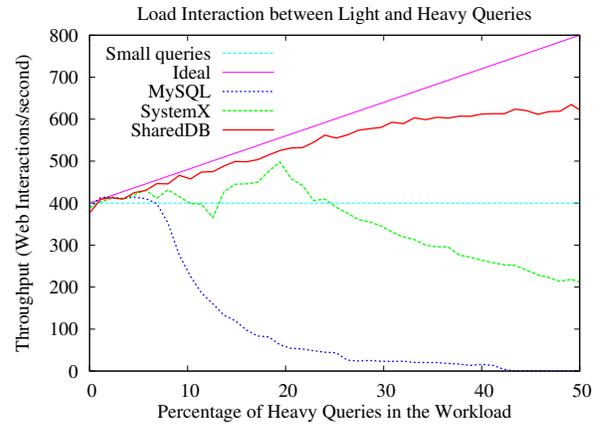

**Figure 11: Load Interaction**

For best sellers queries, the performance of the three systems differs significantly. MySQL's performance is almost linear with the number of queries in the batch. SystemX outperforms MySQL because it is simply the better and more mature system. The best performance, however, was again achieved with SharedDB. Even though SharedDB has a less mature query processor, it outperforms SystemX in this experiment because of sharing.

### 5.7 Load Interaction

In our last experiment, we explored how heavy queries compete with lighter queries for resources and how this resource contention may impact performance. For this reason we used a synthetic workload that is a mixture of the two queries that were analyzed in Section 5.6. A constant load of 400 "search item by title" queries per second was sent to the systems under test. The load of these "search item" queries can easily be sustained by all three systems. In addition to these "search item by title", we submitted an increasing number of "best sellers" queries in this experiment. This way, we were able to study precisely how mixing light and heavy queries affected the performance of the three different systems.

The choice of "search item by title" and "best sellers" queries was not random. In fact, as shown in Figure 6, these two queries have a shared join on *Items* and *Authors* in SharedDB. As a result, these two queries compete for resources in SharedDB, too. Of course, an experiment on resource contention between different kinds of queries would not be fair if SharedDB would run the two different kinds of queries on different cores.

The throughput results for all three systems are shown in Figure 11. At the beginning (the left part of the figure), the load consists of only the 400 "search item" queries and only a few concurrent "best sellers" queries. Such a workload can be sustained by all three systems. Moving to the right (i.e., increasing the number of concurrent "best sellers" queries), MySQL and SystemsX are not longer able to sustain the throughput. In fact, the throughput of MySQL and SystemX drops below 400 so that the presence of "best sellers" queries hurts the execution of the "search item" queries. Obviously, this problem could be fixed by introducing sophisticated load control mechanisms. Figure 11 shows that such load control mechanisms are not needed for SharedDB. The overall throughput increases monotonically; the more concurrent queries, the more sharing and the merrier. Unfortunately, SharedDB is not a perfor-mance panacea either. Starting at about 250 "best sellers" queries per second, SharedDB is not able to handle the full workload either and its throughput diverges from the *ideal* throughput, depicted by the top-most dotted line in Figure 11. Since the concurrent "best sellers" queries have different parameter settings, perfect sharing is not possible and there is a per-query overhead in this experiment which limits the scalability of SharedDB with the number of concurrent queries. Nevertheless, SharedDB scales much better than the other systems, beating SystemX by a factor of 3 in throughput in the extreme case of this experiment. More importantly, SharedDB is robust and makes sure that the processing of heavy queries does not have an impact on the performance of light queries, even if no special load control techniques are applied.

# 6. CONCLUSION

This paper presented SharedDB, a general-purpose relational database system. At the core of SharedDB is a novel query processing model that is based on batching queries and shared computation. SharedDB does not always outperform traditional database techniques that rely on the "query-at-a-time" processing model. The advantages of SharedDB become apparent for high loads with unpredictable mixes of heavy and light queries and updates. In these situations, the performance (i.e., query latency and sustained throughput) of SharedDB is extremely robust without requiring any special tuning knobs, load control, or other adaptive techniques. Our experimental study using the TPC-W benchmark confirmed this robustness along a number of different dimensions: query load, hardware configuration, and query/update diversity. Compared to other related systems that are also based on shared computation, SharedDB wins in terms of generality. For instance, these systems are typically not suitable to process transactional workloads with many small queries and updates.

The SharedDB project is only at its beginning and this paper only presented the main design principles and architecture of SharedDB. The next logical step is to develop a comprehensive query optimizer that automatically generates good global query plans. As part of this work, we will develop a cost model for shared execution in SharedDB. Furthermore, we will extend the SharedDB runtime system in order to make better use of future NUMA machines with possibly hundreds of cores; e.g., integrate parallel joins and advanced partitioning and replication of base data. Another important consideration is to optimize for processor cache locality and integrate, e.g., cache-aware join methods. Currently, SharedDB is based on standard, traditional join methods. Finally, we would like to investigate the benefits of distributing the global query plan across different machines.

*Acknowledgments.* This work was funded by the Enterprise Computing Center of ETH Zurich (www.ecc.ethz.ch). Furthermore, the work of G. Giannikis was supported by a Swiss National Science Foundation grant as part of its Pro-Doc program.